\documentclass[aps,prl,showpacs,twocolumn]{revtex4}

\usepackage{amssymb}
\usepackage{epsfig}
\usepackage{graphicx}
\usepackage{amsmath}
\usepackage{array,color}

\begin{document}

\title{Quantized vortices in a rotating Bose-Einstein condensate with
spatiotemporally modulated interaction}
\author{Deng-Shan Wang}
\author{Shu-Wei Song}
\author{Bo Xiong}
\author{W. M. Liu}
\affiliation{Beijing National Laboratory for Condensed Matter Physics, Institute of
Physics, Chinese Academy of Sciences, Beijing \ 100190, China
}
\date{\today }

\begin{abstract}
We present theoretical analysis and numerical studies of the quantized
vortices in a rotating Bose-Einstein condensate with spatiotemporally
modulated interaction in harmonic and anharmonic potentials, respectively.
The exact quantized vortex and giant vortex solutions are constructed
explicitly by similarity transformation. Their stability behavior has been
examined by numerical simulation, which shows that a new series of stable
vortex states (defined by radial and angular quantum numbers) can be
supported by the spatiotemporally modulated interaction in this system. We
find that there exist stable quantized vortices with large topological
charges in repulsive condensates with spatiotemporally modulated interaction.
We also give an experimental protocol to observe
these vortex states in future experiments.
\end{abstract}

\maketitle

\section{INTRODUCTION}

The investigation of rotating gases or liquids is a central issue in the
theory of superfluidity \cite{Donnelly,Parkins,Pethick,Cooper} since
rotation can lead to the formation of quantized vortices which order into a
vortex array, in close analogy with the behaviour of superfluid helium.
Under conditions of rapid rotation, when the vortex density becomes large,
atomic Bose gases offer the possibility to explore the physics of quantized
vortices in novel parameter regimes. During recent years, there have been
advances in experimental discoveries \cite{Butts,Engels,Schweikhard} of
rotating ultra-cold atomic Bose gases, and these developments have been
reviewed in \cite{Kevrekidis2007}.
\par
Theoretical studies mainly make use of the mean-field Gross-Pitaevskii (GP)
equation to describe the main features of the vortex states \cite{Dalfovo,Aftalion_Ueda}, and several predictions \cite{Aftalion_Ueda} have been shown to agree
with experiments \cite{Rosenbusch}. Some of the important studies were concerned with the
equilibrium properties of a single vortex, including its structures and
dynamics \cite{Aftalion_Ueda}, the critical frequency and the nonlinear
dynamics of vortex lattice formation \cite{Penckwitt}. A multi-quantum
vortex is typically dynamically unstable in harmonically trapped BEC
predicted by several theoretical studies \cite{Pu-PRA-1999,Jackson,JAMH}. The splitting
instability in case of multi-quantum vortices shows that the vortex will
split into single quantum vortices even in the absence of dissipation due to
the peculiar feature of nonlinear dynamics \cite{Kumakura-2007}. However, in
the presence of a plug potential \cite{Simula-PRA-2002} or an anharmonic
trapping potential \cite{Lundh-PRA-2002,Ketterle2001}, various studies have addressed
different means to stabilize multi-quantum vortices in rotating BEC. For
example, when the confining potential is steeper than harmonic potential in
the plane perpendicular to the axis of rotation, multi-quantum vortices are
energetically favorable if the interaction is weak enough. For stronger
interactions, the multiply quantized vortices break up into arrays of
several vortices. In addition, interestingly, stable multi-quantum vortices
have also been found to exist in two-component BEC \cite%
{Ruostekoski-PRA-2004}, which can be adjusted near Feshbach resonance through spatial inhomogeneous external magnetic field $B,$  i.e. $B=B(x)$.

Mathematically, the GP equation, to be written explicitly, is an equation of
nonlinear Schr\textrm{\"{o}}dinger type \cite%
{Sulem,Belmonte-Beitia1,Wang,Belmonte-Beitia2,Wu,Yan_Konotop,Kasamatsu}.
This equation has been studied extensively both in the physical and
mathematical literature, since they provide a universal model for a study of
the dynamics of the envelope waves. One of the distinctive features of the
equation as it appears in BEC problems is the presence of an external
trapping potential, which essentially affects the elementary excitation
spectrum. Most properties of the BEC are significantly affected by the
interatomic interaction, which can be characterized by the $s$-wave
scattering length \cite{Timmermans}. Recent experiments have demonstrated
that both amplitude and sign of the scattering length can be modified by
utilizing the Feshbach resonance \cite{Roberts}. This technique provides a
very promising method for the manipulation of atomic matter waves and the
nonlinear excitations in BEC by tuning the interatomic interaction. By using
this technique, one can study atomic matter waves and the nonlinear
excitations in BEC for the case of the GP equations with the time- and
space-dependent nonlinearity coefficients \cite%
{Belmonte-Beitia1,Wang,Belmonte-Beitia2,Wu}.
\par
Motivated by stabilizing multiple vortex states and understanding the
behavior of nonlinear excitation in physical systems, we perform theoretical
analysis and numerical studies of the quantized vortices in a rotating BEC
with spatiotemporally modulated interaction in harmonic and anharmonic
potentials, respectively. Compared with the former work on quantized
vortices, we find that a new series of exact single and multiple vortex states
(defined by radial and angular quantum numbers) can be supported by the
spatiotemporally modulated interaction in a rotating BEC. In particular, our
results have provided a very promising method for stabilizing the vortex
having very large topological charge $S\geq 2$, which has been conjectured
unstable \cite{Mihalache} by tuning the external potential and nonlinear
interaction simultaneously in time.

\section{The Theoretical Model and Exact Vortex Solutions}

At zero temperature, the quantum and thermal fluctuations are negligible so
that a BEC trapped in an external potential can be described by a
\textquotedblleft macroscopic wave function" $\Psi(\mathbf{r},t)$ obeying
the GP equation. In the rotating frame with rotating frequency $\Omega_0$
around the $z$-axis, the GP equation in cylindrical coordinate reads \cite%
{Ueda}

\begin{equation}
i\hbar\frac{\partial \Psi}{\partial t} \!=\! ( - \frac{\hbar^2}{2m} \nabla^2
\!- \frac{\hbar^2}{2m}\frac{\partial^2}{\partial z^2}+\! V_{ ext} \!+\!
G|\Psi|^2) \Psi +i\hbar\Omega_0 \,\frac{\partial \Psi }{\partial \theta },
\notag  \label{GP}
\end{equation}
where $\nabla^2=\partial^2/\partial x^2+\partial^2/\partial
y^2=\partial^2/\partial r^2+1/r\times\partial/\partial r
+1/r^2\partial^2/\partial \theta^2$ with $r^2=x^2+y^2,$ $m$ is the atom
mass, $\theta $ is the azimuthal angel, the wave function is normalized by
the total particle number $N= \int d\mathbf{r}|\Psi|^2,$ $V_{ ext}$ is an
external trapping potential, and $G=4\pi\hbar^2a(r,t)/m$ represents the
strength of interatomic interaction characterized by the $s$-wave scattering
length $a(r,t)$, which can be adjusted experimentally by an inhomogeneous external magnetic field $B=B(x,y,t)$ in the vicinity of a Feshbach resonance \cite{Roberts}.
The trapping potential can be assumed to be $V_{ ext}=m(\omega_r^2
r^2+\omega_z^2z^2)/2$, where $\omega_{r}$ and $\omega_{z}$ are the
confinement frequencies in the radial and axial directions, respectively,
and in particular, the radial confinement frequency $\omega _{r}$ is assumed
to be time-dependent as in \cite{Belmonte-Beitia2,Saito_Ueda}. In the
following, we consider the atoms in
the $|F=1, m_F=1\rangle$ hyperfine state of $^{7}$Li and $|F=1, m_F=1\rangle$ hyperfine state of $^{87}$Rb
trapped in a very thin disc-shaped potential,
i.e., the trapping potential in the radial direction is much weaker than
that in the axial direction as $\omega _{r}(t)/\omega _{z}\ll 1$, such that
the motion of atoms in the $z$ direction is essentially frozen to the ground
state $\varphi(z)$ of the axial harmonic trapping potential.

Then we can separate the wave function as $\Psi(\mathbf{r},t)=\psi(x, y,
t)\varphi(z)$ to derive the 2D GP equation
\begin{equation}
i\hbar\frac{\partial \psi}{\partial t} = - \frac{\hbar^2}{2m} \nabla^2\psi
\!+\! \frac{m}{2}\omega_r^2r^2\psi \!+\! G \eta|\psi|^2\psi +i\hbar\Omega_0
\,\frac{\partial \psi }{\partial \theta },  \label{GP0}
\end{equation}
with $\eta={\int dz| \varphi(z)|^4}/{\int dz| \varphi(z )|^2}. $ Introducing
the scales characterizing the trapping potential, the length, time, and wave
function are scaled as
\begin{equation}
x=a_h\tilde{x}, ~~y=a_h\tilde{y},~~t={\tilde{t}}/{\omega_z},~~ \psi={\tilde{%
\psi}}/{a_h\sqrt{4\pi a_0\eta}},  \notag
\end{equation}
respectively, with $a_h=(\hbar /m\omega _{z})^{1/2}$ and $a_0$ is a constant
length chosen to measure the $s$-wave scattering length. After the tilde is
omitted, the 2D GP equation (\ref{GP0}) is reduced to a dimensionless form
as
\begin{equation}
i\frac{\partial \psi }{\partial t}=-\frac{1}{2}\,\nabla ^{2}\psi +V(r,t)\psi
+g(r,t)\left\vert \psi \right\vert ^{2}\psi +i\Omega \,\frac{\partial \psi }{%
\partial \theta },  \label{GP1}
\end{equation}
where the interaction strength $g(r,t)=a(r,t)/a_0,$ $\Omega=\Omega_0/%
\omega_z $ and the radial trapping potential can be written as $%
V(r,t)=\omega ^{2}(t)r^{2}/2$ with $\omega (t)=\omega _{r}(t)/\omega _{z}.$
In what follows, we consider not only the harmonic potential like this but
also an anharmonic potential.

In order to find the exact vortex solutions to Eq. (\ref{GP1}) with
spatiotemporally modulated interaction, we first assume its exact solution
as
\begin{equation}
\psi (r,\theta ,t)={e^{iS\theta +i\phi (r,t)}}\rho (r,t)\,U\left[ R(r,t)%
\right] ,  \label{sol_1}
\end{equation}%
here $S$ is the topological charge related to the angular momentum of the
condensate, $\rho(r,t)$ denotes the amplitude of wave function and $R(r,t)$ is a intermediate variable reflecting the changes of main wave function $U$. Substituting Eq. (\ref{sol_1}) into (\ref{GP1}) and furthermore,
letting $U[R(r,t)]$ satisfy
\begin{equation}
d^{2}U/dR^{2}+\mu _{0}U+\mu _{1}U^{3}=0,  \label{sol_2}
\end{equation}%
where $\mu _{0}$ and $\mu _{1}$ are real constants, we can get a set of
partial differential equations (PDEs) about $\rho (r,t),R(r,t),\phi
(r,t),V(r,t)$ and $g(r,t)$ as
\begin{gather}
R_{{t}}+\phi _{{r}}R_{{r}}=0,  \notag \\
2\,g{\rho }^{2}+\mu _{{1}} \,{R_{{r}}}^{2}=0,  \notag \\
\rho \,R_{{r}}+2\,r\rho _{{r}}R_{{r}}+r\rho \,R_{{rr}}=0,  \notag \\
r\rho \phi _{{rr}}+2\,r\phi _{{r}}\rho _{{r}}+\rho \phi _{{r}}+2\,r\rho _{{t}%
}=0,  \label{PDE-1} \\
{\frac {\rho_{{rr}}}{\rho}}-2\,\phi_{{t}}-{\phi_{{r}}}^{2}-{\frac {{\ S}^{2}%
}{{r}^{2}}}+2\,\Omega\,S+{\frac {\rho_{{r}}}{\rho\,r}}-2\,V-\mu_ {{0}}{R_{{r}%
}}^{2}=0,  \notag
\end{gather}%
here subscripts $r$ and $t$ mean the derivative of function with respect to $%
r$ and $t$. If letting $V(r,t)={\omega (t)}^{2}{r}^{2}/2-\mu _{0}R_{r}^{2}/2$
($\mu _{0}=0$ corresponds to harmonic potential) and $\phi (r,t)=f_{{1}}{r}%
^{2}+f_{{2}}$ ($f_{{1}}$ and $f_{{2}}$ are time-dependent functions, and $f_1$ is frequency chirp and $f_2$ is phase), then
solving the set of PDEs (\ref{PDE-1}) we get
\begin{equation}
\rho (r,t)={e^{-2\,\int \!f_{{1}}{dt}}}\Theta (r{e^{-2\int f_{{1}}{dt}}}),
\label{rau}
\end{equation}%
and
\begin{equation}
R(r,t)=\int_{0}^{r{e^{-2\int f_{{1}}{dt}}}}1/[\Theta ^{2}(\tau )\tau ]d\tau ,
\label{R}
\end{equation}%
and%
\begin{equation}
g(r,t)=-\mu _{1}R_{r}^{2}/2\rho ^{2},  \label{G-interaction}
\end{equation}%
where $\mu_1$ is a parameter controlling the sign of interaction parameter $g(r,t)$, $\Theta (\tau )$ is defined by Whittaker M and W functions \cite%
{Abramowitz}, i.e. $\Theta (\tau )=[c_{{1}}{M}({\lambda _{{1}}/4\lambda _{{2}%
}},S/2,\lambda _{{2}}\tau ^{2})+c_{{2}}{W}({\lambda _{{1}}/4\lambda _{{2}}}%
,S/2,\lambda _{{2}}\tau ^{2})]/\tau ,$ where $\lambda _{1},\lambda
_{2},c_{1},c_{2}$ are nonzero constants and $c_{1}c_{2}>0$. In particular,
the above $f_{1}$ and $f_{2}$ satisfy the following two ordinary
differential equations
\begin{gather}
2\,\Omega \,S-\lambda _{{1}}{e^{-4\,\int \!f_{{1}}{dt}}}-2{df_{{2}}}/{dt}=0,
\notag \\
\omega ^{2}(t)+4\,f_{{1}}^{2}+2\,df_{{1}}/dt-{\lambda^{2} _{{2}}}{%
e^{-8\,\int \!f_{{1}}{dt}}}=0.  \label{sol_3}
\end{gather}

\begin{figure}[tbp]
\includegraphics[width=9cm]{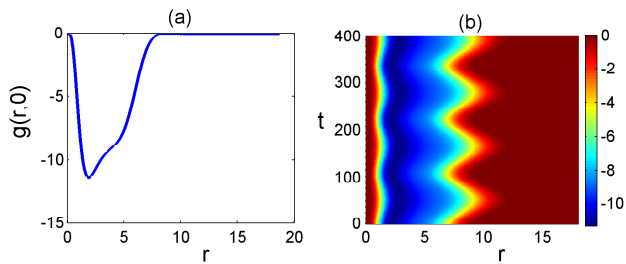}
\caption{{\protect\small (Color online) The spatiotemporally dependent
interaction parameter $g(r,t)$ in Eq. (\protect\ref{G-interaction}) for
parameter $\Omega =0.7, \protect\lambda_{1}=4, \protect\lambda_{2}=2,
c_{1}=c_{2}=1, S=1, \mu_1=1000$ and $\protect\omega(t)=0.028.$ (a) The radial
structure of $g(r,t)$ at $t=0$. (b) The spatiotemporal structure of $g(r,t)$%
.}}
\label{Fig_0}
\end{figure}

When the parameter $\mu _{0}=0$, the external potential is just harmonic
form $V(r,t)={\omega }^{2}(t){r}^{2}/2$, and we get explicit solution of Eq.
(\ref{sol_2}) as $U(R)=\nu _{1}\mathrm{cn}(\nu _{1}\,R+\nu _{0},\sqrt{2}/{2}%
)/\sqrt{\mu _{{1}}}$, where $\nu _{0}$ and $\nu _{1}$ are arbitrary
constants and $\mu _{1}>0$. So the exact vortex solution to Eq. (\ref{GP1})
is
\begin{equation}
\psi =\frac{{{\nu _{1}}}\,}{\sqrt{\mu _{{1}}}}{{\rho }e^{i\left( S\theta +f_{%
{1}}{r}^{2}+f_{{2}}\right) }}{\ }\mathrm{cn}(\nu _{1}\,R+\nu _{0},\sqrt{2}/{2%
}),  \label{sol_4}
\end{equation}%
where functions $\rho $ and $R$ are given above. Here $\mathrm{cn}$ and $%
\mathrm{sn}$ (below) are Jacobi elliptic functions. When imposing the
boundary conditions for vortex solution as $\lim_{|r|\rightarrow 0}\psi
(r,\theta ,t)=\lim_{|r|\rightarrow \infty }\psi (r,\theta ,t)=0,$ we can get
$\nu _{0}=K(\sqrt{2}/{2}),\nu _{1}=2nK(\sqrt{2}/{2})/R(+\infty ,0)$, where $%
K(s)=\int_{0}^{\pi /2}[1-s^{2}\sin ^{2}\tau ]^{-1/2}d\tau $ is the first
kind elliptic integral and $n$ is a nonnegative integer. In this case, since
$\mu _{{1}}$ should be positive and thus the interaction strength $g(r,t)$
is negative corresponding to the condensates consisting of $^{85}$Rb \cite%
{Cornish} or $^{7}$Li atoms \cite{Bradley,Gerton} experimentally. The structures of the interaction parameter $g(r,t)$ with respect to radial coordinate $r$ and time $t$ are demonstrated in Fig. \ref{Fig_0}. It is observed that the interaction parameter is inverse Gaussian in $r$ and periodic in $t$.

\begin{figure}[tbp]
\includegraphics[width=9.5cm]{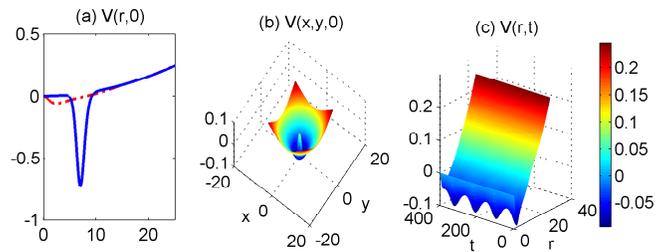}
\caption{{\protect\small (Color online) The shapes of the anharmonic
potential $V(r,t)={\protect\omega }^{2}(t){r}^{2}/2-\protect\mu %
_{0}R_{r}^{2}/2$. (a) The radial structures at $t=0$ with parameters $S=1,
\protect\delta=7.4$ (dashed line) and $S=5, \protect\delta=1172.4$ (solid
line), respectively. (b) The spacial structure at $t=0$ with parameter $S=1, \protect%
\delta=7.4$. (c) The spatiotemporal structure with parameter $S=1, \protect%
\delta=7.4$. The other parameters are $\Omega =0.7, \protect\lambda_{1}=4,
\protect\lambda_{2}=2, c_{1}=c_{2}=1, \mu_0=62.2$ and $\protect\omega(t)=0.028.$ }}
\label{Fig_1}
\end{figure}

\par
When the parameter $\mu _{0}\neq 0,$ the external potential becomes $V(r,t)={%
\omega }^{2}(t){r}^{2}/2-\mu _{0}R_{r}^{2}/2$ (an anharmonic potential) as shown in Fig. \ref{Fig_1},
where there is a convex hull in the center of the harmonic potential, and the anharmonic potential is periodic in time $t$. We get
the exact solution of Eq. (\ref{sol_2}) as $U(R)=\sqrt{{{2\,({\delta }%
^{2}-\mu _{{0}}})/{\ \mu _{{1}}}}}\mathrm{sn}\left( \delta R,{\ \sqrt{\mu _{{%
0}}/\delta ^{2}-1}}\right) ,$ where $\mu _{0}/2<\delta ^{2}<\mu _{0}$ and $%
\mu _{1}<0$. So the exact vortex solution to Eq. (\ref{GP1}) is%
\begin{equation}
\psi =\sqrt{\frac{{{2\,({\delta }^{2}-\mu _{{0}}})}}{{{\mu _{{1}}}}}}\rho {%
e^{i\left( S\theta +f_{{1}}{r}^{2}+f_{{2}}\right) }}\mathrm{sn}\left( \delta
R,{\ \sqrt{\mu _{{0}}/\delta ^{2}-1}}\right) ,  \label{sol_5}
\end{equation}
where functions $\rho $ and $R$ are given above. When imposing the boundary
conditions for vortex solution as $\lim_{|r|\rightarrow 0}\psi (r,\theta
,t)=\lim_{|r|\rightarrow \infty }\psi (r,\theta ,t)=0$, we can get $\delta
=2nK({\ \sqrt{\mu _{{0}}/\delta^{2}-1}})/R(\infty ,0).$ In this case, the
parameter $\mu _{{1}}$ should be negative and thus the interaction strength $%
g(r,t)$ is positive corresponding to the condensates consisting of $^{87}$Rb
\cite{Anderson} or $^{23}$Na atoms \cite{Moerdijk} experimentally.

\begin{figure}[tbp]
\includegraphics[width=9.0cm]{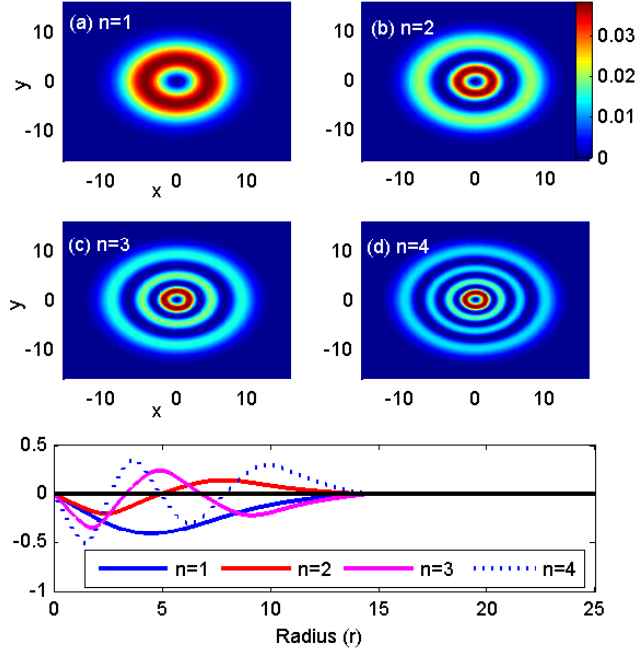}
\caption{{\protect\small (Color online) The density distributions $|\protect%
\psi (x,y,0)|^{2}$ (a)-(d), and the corresponding radial wave profiles
(bottom) for the vortex solution (\protect\ref{sol_4}) of the attractive
rotating BEC for topological charge $S=1$ and various radial quantum
numbers. The parameters are $\Omega =0.7, \protect\mu _{1}=1000,\protect%
\lambda _{1}=4,\protect\lambda _{2}=2,c_{1}=c_{2}=1, \protect\epsilon =0$
and $\protect\omega_0=0.028.$ }}
\label{Fig_2}
\end{figure}

Seen from the exact vortex solutions (\ref{sol_4}) and (\ref{sol_5}), there
exists two class of vortex states (distinguishing them with two quantum
numbers which are radial node $n$ and topological charge $S$, also called
angular momentum quantum number) corresponding to the harmonic potential $%
\left( \mu _{0}=0\right) $ and anharmonic potential $\left( \mu _{0}\neq
0\right) $ in attractive and repulsive BECs, respectively. In the following,
we will first examine the structures of these exact vortex solutions and
then study the dynamic properties and stability of these vortex states under
different situations.

\section{Structures of Vortex States}

The structures of the exact vortex solutions (\ref{sol_4}) and (\ref{sol_5})
can be controlled by modulating the frequency of the trapping potential and
the spatiotemporal inhomogeneous $s$-wave scattering length as seen from Eq.
(\ref{sol_3}). Taking into account the feasibility of the experiment, we
only consider the case of harmonic potential $\left( \mu _{0}=0\right) $
which corresponds to the attractive BEC as explained above.
\par
In real experiment, we assume an attractive $^{7}$Li condensate in the internal atomic state $|F=1, m_F=1\rangle$ \cite{Bradley,Gerton}
trapped in an axis-symmetric disk-shaped potential, where the
axial confinement energy $\hbar \omega _{z}$ is much larger than the radial
confinement and interaction energies, and the radial frequency of the trap
is time-dependent which can be written as%
\begin{equation}
\omega (t)=\omega _{r}(t)/\omega _{z}=\omega _{0}+\epsilon \cos (\omega
_{1}t),  \label{omiga_1}
\end{equation}%
with $0\leq \epsilon <\omega _{0}$. For $\epsilon =0,$ the radial frequency
of the trap is time-independent and here, we choose the time-independent
part of radial frequency $\omega _{r}=(2\pi) \times 18 $ Hz and axial
frequency $\omega _{z}=(2\pi)\times 628 $ Hz as in \cite{Theocharis}, so $%
\omega (t)=\omega _{0}=0.028$.

\begin{figure}[tbp]
\includegraphics[width=9.5cm]{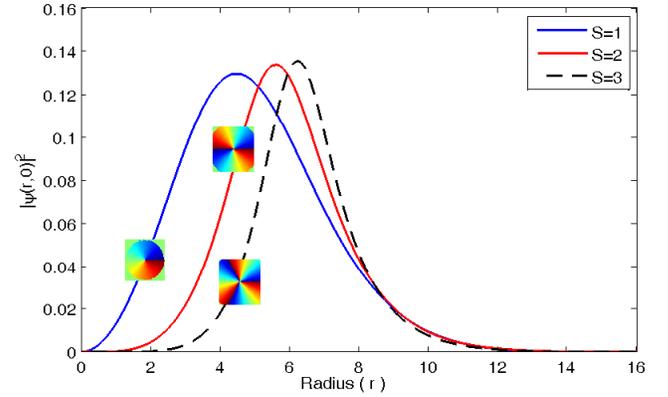}
\caption{{\protect\small (Color online) The radial structures of the density
distributions and phase diagrams for the vortex solution (\protect\ref{sol_4}%
) of the attractive rotating BEC with radial quantum number $n=1$ and
different topological charges at $t=0$. The insets are the corresponding
phase diagrams and the other parameters are the same as Fig. 3. }}
\label{Fig_3}
\end{figure}

\begin{figure}[tbp]
\includegraphics[width=9.5cm]{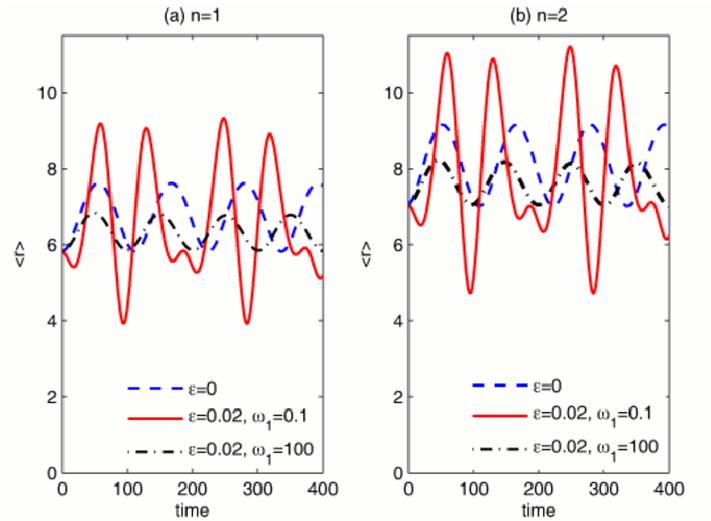}
\caption{{\protect\small (Color online) Time evolution of the monopole
moment $\langle r \rangle$ in the attractive rotating BEC for different
frequencies of the harmonic potential. (a) The radial quantum number is $n=1$
and (b) $n=2$. In both figures, the topological charge is $S=1$ and the
frequency of the trap is $\protect\omega (t)=\protect\omega _{0}+\protect%
\epsilon \cos ( \protect\omega _{1}t)$ with parameters $\Omega =0.7, \protect%
\omega _{0}=0.028, \protect\lambda _{1}=4,\protect\lambda _{2}=2, \protect\mu%
_{1}=1000,c_{1}=c_{2}=1.$ }}
\label{Fig_4}
\end{figure}

As shown in Fig. \ref{Fig_2}, we demonstrates the density distributions for
different radial quantum number $n$ with fixed topological charge $S=1$ at $%
t=0$, which is based on the exact vortex solution (\ref{sol_4}). The Fig. %
\ref{Fig_2}(a) corresponding to $n=1$ is a lowest energy state and Figs. \ref%
{Fig_2}(b)-\ref{Fig_2}(d) corresponding to $n=2,3,4$ are three excited
states. In the Fig. \ref{Fig_2}(e), we show the radial wave profiles of the
exact vortex solution (\ref{sol_4}) at $t=0$. It is clear to see that the
number of ring structure of vortex solution increases by one with changing
the radial quantum number $n$ by one, which is similar to the quantum states
of harmonic oscillator.

One of the interesting properties for the exact vortex solution (\ref{sol_4}%
) is shown in Fig. \ref{Fig_3} by choosing different topological charge $S$
with fixed radial quantum number $n=1$. We can see that the density profiles
of the vortex states become more and more localized with increasing the
topological charge $S$ due to the larger angular momentum for the higher
topological charge $S$. Moreover, vortex expands outwards with the
increasing of the topological charges and so will obtain the larger angular
momentum.

Another interesting aspect of the condensate is to study the monopole moment
\cite{Saito_Ueda,Ueda_Liu} defined by $\left\langle r\right\rangle =\int r|\psi |^{2}dr$ which can be
directly compared with experiments in BEC. In Fig. \ref{Fig_4}, we show the
time evolution of the monopole moment for different oscillation frequencies $%
\omega _{1}$, amplitude $\epsilon $ in (\ref{omiga_1}), and different radial
quantum numbers $n=1$, $2$ with fixed topological charge $S=1$. It is seen
that the monopole moment represents regular oscillation following the
oscillating frequency of the trap when $\omega _{1}$ is large, but irregular
oscillation when $\omega _{1}$ is small which can be observed
experimentally. The amplitude and center position of oscillation of monopole
moment for the case $n=2$ are a little larger than that for the case $n=1$,
but their periods are the same. In particular, at $\epsilon =0$ i.e.
time-independent trapping potential, the oscillation of the monopole moment
is only determined by the spatiotemporally nonlinear interaction which also
represents regular behavior in our studied case.

\begin{figure}[tbp]
\includegraphics[width=9.5cm]{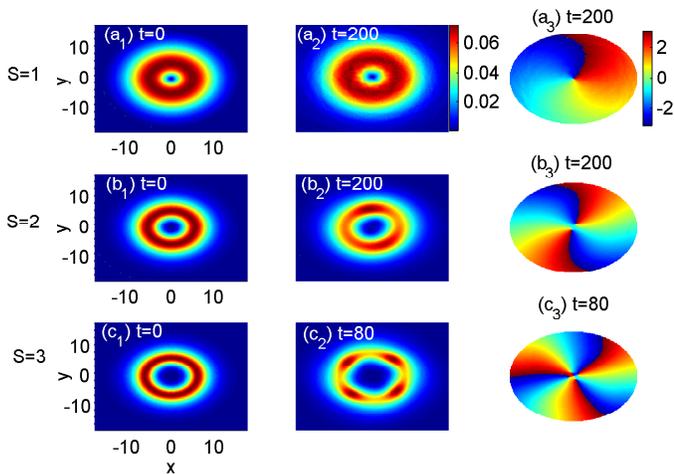}
\caption{{\protect\small (Color online) Time evolution of the density
distributions $|\protect\psi (x,y,0)|^{2}$ and phase diagrams for the vortex
solution (\protect\ref{sol_4}) of the attractive rotating BEC for radial
quantum number $n=1$. (a) Stable vortex for topological charge $S=1.$
(b)-(c) Unstable vortex for $S=2$ and $3$, respectively. For all cases, the
domain is $(x,y)\in [-15,15]\times [-15,15]$. The other parameters are $%
\Omega =0.7, \protect\lambda _{1}=4, \protect\lambda _{2}=2, \protect\mu%
_{1}=1000, c_{1}=c_{2}=1, \protect\epsilon=0$ and $\protect\omega_0=0.028$. }
}
\label{Fig_5}
\end{figure}

\section{ Stability Analysis}

\begin{figure}[tbp]
\includegraphics[width=10cm]{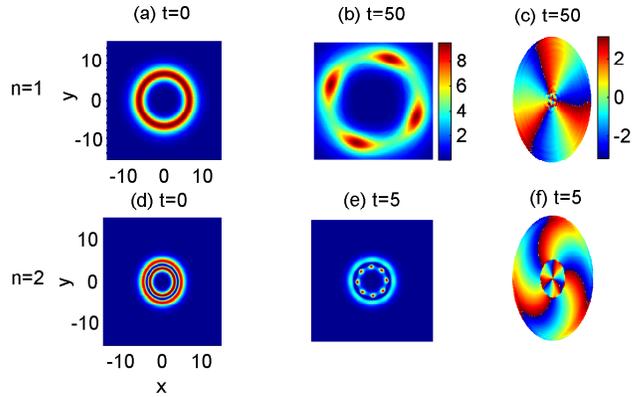}
\caption{{\protect\small (Color online) The dynamic instability of the
vortex solution (\protect\ref{sol_4}) of the attractive rotating BEC with
time-dependent harmonic potential for topological charge $S=3$ and two
different radial quantum numbers. The parameters are $\Omega=0.7, \protect%
\lambda_{1}=10, \protect\lambda_{2}=5,\protect\mu_{1}=1000,c_{1}=c_{2}=1,
\protect\omega _{1}=1, \protect\epsilon =0.02$ and $\protect\omega%
_{0}=0.028. $ }}
\label{Fig_6}
\end{figure}

It has been shown that attractive Bose condensates like $^{85}$Rb and $^{7}$%
Li become mechanically unstable and collectively collapse \cite%
{Roberts,Gerton} when the number of atoms in the condensate exceeds critical
value $N_{c}$. So it is important to produce the stable states in attractive
Bose condensates. Saito and Ueda \cite{Saito_Ueda} have demonstrated that a
matter-wave bright soliton can be stabilized in 2D free space by causing the
strength of interactions to oscillate rapidly between repulsive and
attractive by using, e.g., Feshbach resonance \cite{Roberts}. In previous
work, we \cite{Wang} have found an exact stable localized nonlinear matter
wave in quasi-2D BEC with spatially modulated nonlinearity in harmonic
potential. In this section, we investigate the dynamical stability of the
exact vortex solutions (\ref{sol_4}) and (\ref{sol_5}) by numerical
simulation of Eq. (\ref{GP1}). We show that only some types of the stable
vortices (defined by radial and angular quantum numbers) can be supported by
the spatiotemporally modulated interaction in this system.

\begin{figure}[tbp]
\includegraphics[width=9cm]{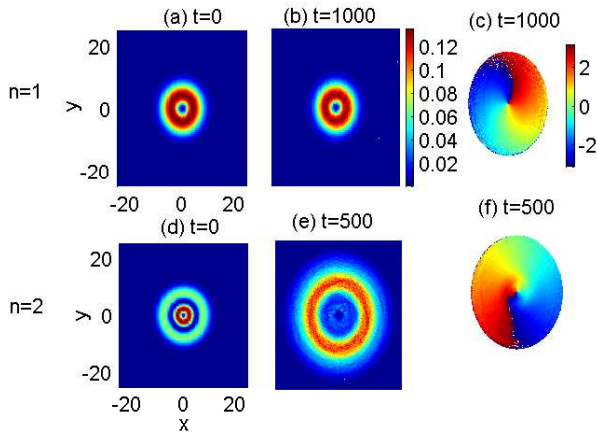}
\caption{{\protect\small (Color online) Time evolution of the density
distributions $|\protect\psi (x,y,0)|^{2}$ and phase diagrams for the vortex
solution (\protect\ref{sol_5}) of the repulsive rotating BEC for topological
charge $S=1$ and radial quantum numbers $n=1$ and $2,$ respectively, with an
initial Gaussian noise of level 0.5\%. Here the domain is $(x,y)\in
[-25,25]\times [-25,25],$ and the parameters are $\protect\delta=7.4$ (top)
and $\protect\delta=14.9$ (bottom). The other parameters are $\Omega =0.7,
\protect\lambda_{1}=4, \protect\lambda_{2}=2, c_{1}=c_{2}=1, \protect\mu%
_{1}=-10, \protect\epsilon=0$ and $\protect\omega_0=0.028.$ }}
\label{Fig_7}
\end{figure}

In order to elucidate the dynamical stability of the exact vortex solutions
proposed in Section II, we conduct numerical experiments by solving Eq. (\ref%
{GP1}) and take the exact vortex solutions (\ref{sol_4}) and (\ref{sol_5})
at $t=0$ as initial data. To begin with, we consider the attractive rotating
BEC with harmonic potential at $\epsilon =0$ in (\ref{omiga_1}), which has
exact vortex solution (\ref{sol_4}). In Fig. \ref{Fig_5}, we show the
density evolutions and phase diagrams of vortex solution (\ref{sol_4}) as
initial condition with radial quantum number $n=1$ and different topological
charge $S$ or angular momentum quantum numbers based on numerical simulation
of Eq. (\ref{GP1}). It is shown that only when topological charge $S=1$,
vortex solution (\ref{sol_4}) is stable against perturbation with an initial
Gaussian noise of level 0.5\%, but for topological charge $S\geq 2$, giant
vortex solution (\ref{sol_4}) will be unstable and split into single charge
vortices and so destruct the ring structures.

When the harmonic trap is time-dependent which corresponds to $\epsilon \neq
0$ and $\omega _{1}\neq 0$ in (\ref{omiga_1}), Fig. \ref{Fig_6} shows the
unstable dynamics and phase diagrams of the giant vortex solution (\ref%
{sol_4}) with $S=3$ and two different radial quantum numbers $n=1,2$ for the
attractive rotating BEC. It is observed that the time-dependent frequency of
trap affects the dynamics of the vortex significantly.

\par
\begin{figure}[tbp]
\includegraphics[width=9cm]{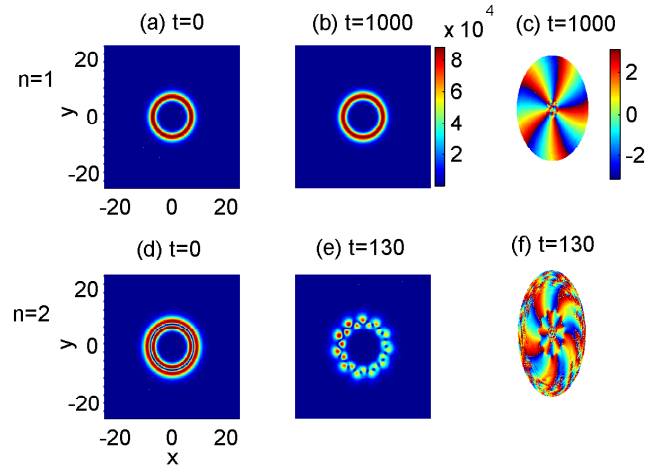}
\caption{{\protect\small (Color online) Time evolution of the density
distributions $|\protect\psi (x,y,0)|^{2}$ and phase diagrams for the vortex
solution (\protect\ref{sol_5}) of the repulsive rotating BEC for topological
charge $S=5$ and radial quantum numbers $n=1$ and $2,$ respectively, with an
initial Gaussian noise of level 0.5\%. Here the domain is $(x,y)\in
[-25,25]\times [-25,25],$ and the parameters are $\protect\delta=1172.4$
(top) and $\protect\delta=2344.9$ (bottom). The other parameters are the
same as Fig. 8. }}
\label{Fig_8}
\end{figure}

Next we consider the repulsive rotating BEC in anharmonic potential $V(r,t)={%
\omega }^{2}(t){r}^{2}/2-\mu _{0}R_{r}^{2}/2 $ with $\mu _{0}\neq 0 $ shown
in Fig. \ref{Fig_1}, which has exact vortex solution (\ref{sol_5}). In Fig. %
\ref{Fig_7} and \ref{Fig_8}, we demonstrate the density evolutions and phase
diagrams of vortex solution (\ref{sol_5}) as initial condition with
different radial quantum numbers $n=1,2$ and fixed angular momentum quantum
numbers $S=1 $ and $S=5$, respectively. It is very interesting to note that
when the radial quantum number $n=1,$ the exact vortex solution (\ref{sol_5}%
) is always stable even for very large topological charge $S=5,$ which is
very different from the attractive rotating BEC with harmonic potential
where a stable region for vortex solution (\ref{sol_4}) was found only for $%
S=1$ as shown in Fig. \ref{Fig_5}. Our results have provided a very
promising method for stabilizing the giant vortex having very large
topological charge $S\geq 2$ which has been conjectured unstable \cite{Mihalache}
by tuning the external potential and nonlinear interaction simultaneously in
time. Numerical simulation shows that for the radial quantum number $n>1$, the vortex solution (\ref{sol_5})
is always unstable for any topological charge $S$.

\par
Finally, we investigate the effect of the slightly asymmetrical potential to the stability of quantized vortices.
To do so, we take the asymmetrical external trap as $V(r,t)={
\omega }^{2}(t)[(1+\epsilon_x)x^2+(1+\epsilon_y)y^2]/2-\mu _{0}R_{r}^{2}/2 $, where parameters $\epsilon_x$ and
$\epsilon_y$ describe small deviations of the trap from the axisymmetry. The ENS group \cite{Madison2000} stirred a BEC of $^{87}$Rb confined in this kind of magnetic trap using a focused laser beam.
Fig. 10 shows the evolutions of density
profiles and phase diagrams of the vortex
solution (11) at $t=1000$ for the inhomogeneous repulsive rotating BEC in this slightly asymmetrical anharmonic potential with parameters $\epsilon_x=0.02$ and $\epsilon_y=0.03$. Here the quantum number $n=1$ and topological
charges $S=1$ and $5$, respectively, and the level of the
initial Gaussian noise is still 0.5\%. It is seen that the quantized vortices with $S=1$ and $5$ in slightly asymmetrical anharmonic potential are still stable at $t=1000$.
\par
We now give an experimental protocol to observe the above vortex states in future experiments.
For the attractive interactions, we take $^{7}$Li condensate in internal atomic state $|F=1, m_F=1\rangle$ \cite{Bradley,Gerton}, containing about $6.55\times10^4$
atoms, confined in a pancake-shaped trap with
radial frequency $\omega _{r}=(2\pi) \times 18 $ Hz and axial
frequency $\omega _{z}=(2\pi)\times 628 $ Hz \cite{Theocharis}. Experimentally, this trap can be determined by
combination of spectroscopic observations, direct magnetic field
measurement, and the observed spatial cylindrical symmetry of the
trapped atom cloud \cite{Rychtarik}. For the repulsive interactions, we take $^{87}$Rb condensate in internal atomic state $|F=1, m_F=1\rangle$ \cite{Anderson}, containing about $8 \times 10^5$
atoms, confined in an anharmonic potential which is a pancake-shaped harmonic trap with
radial frequency $\omega _{r}=(2\pi) \times 18 $ Hz and axial
frequency $\omega _{z}=(2\pi)\times 628 $ Hz \cite{Theocharis} plus a convex hull, see Fig. 2.
The key step is how to realize the
spatiotemporal variation of the scattering length. Near the Feshbach
resonance \cite{Roberts}, the scattering
length $a_s(B)$ varies dispersively as a function of magnetic field
$B,$ i.e. $a_s(B)=\tilde{a}[1-\Delta/(B-B_0)],$ with $\tilde{a}$ being the
asymptotic value of the scattering length far from the resonance,
$B_0$ being the resonant value of the magnetic field, and $\Delta$ being the width of the resonance at $B=B_0$. For the magnetic field in $z$
direction with gradient $\alpha$ along $x$-$y$ direction, we have
$\vec{B}=[B_0+\alpha B_1(x,y,t)]\vec{e_z}$. In this case, the
scattering length is dependent on $x, y$ and time $t$. So in real experiments, we can use Feshbach
resonance technique to realize spatiotemporal variation of interaction parameters shown in Fig. 1.
Finally, in order to observe the density distributions in
Figs. 3 and 6-9 clearly in experiment, the atoms should be
evaporatively cooled to low temperatures, say in the range of 10 to
100 $nK$.

\begin{figure}[tbp]
\includegraphics[width=9cm]{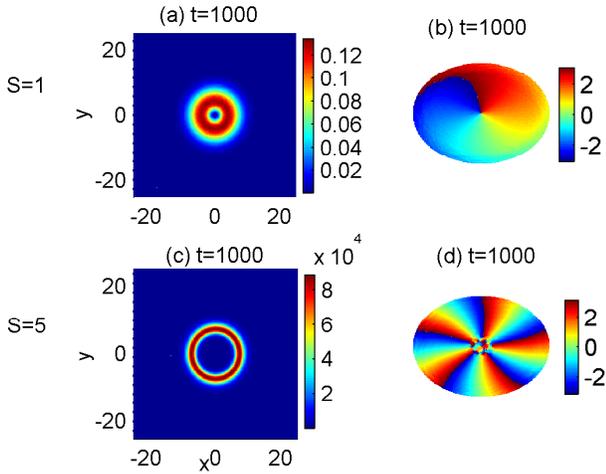}
\caption{{\protect\small (Color online) Density
profiles and phase diagrams for the vortex
solution (\protect\ref{sol_5}) at $t=1000$ for the repulsive inhomogeneous rotating BEC in slightly asymmetrical anharmonic potential $V(r,t)={%
\omega }^{2}(t)[(1+\epsilon_x)x^2+(1+\epsilon_y)y^2]/2-\mu _{0}R_{r}^{2}/2 $ with $\epsilon_x=0.02$ and $\epsilon_y=0.03$.
 Here the quantum numbers $n=1$ and topological
charges $S=1$ and $5$, respectively, and the level of the
initial Gaussian noise is still 0.5\%. The parameters are $\protect\delta=7.4$
(top) and $\protect\delta=1172.4$ (bottom). The other parameters are the
same as Figs. 8 and 9. }}
\label{Fig_10}
\end{figure}

\section{CONCLUSIONS}

In conclusion, we have investigated the quantized vortices in a rotating BEC
with spatiotemporally modulated interaction in harmonic and anharmonic
potentials, respectively. Two families of exact vortex solutions for the 2D
GP equation are constructed explicitly by similarity transformation. It is
interesting to see that a new series of stable giant vortex states with
topological charge $S\geq 2$ can be supported by tuning the external potential
and the spatiotemporally modulated interaction in this system. We hope that
this paper will stimulate further research on quantized vortices and help to
understand the behavior of nonlinear excitation in physical systems with
spatiotemporally modulated interaction.\newline

\textbf{Acknowledgments}

This work was supported by NSFC under grants Nos. 10874235,
10934010, 60978019 and 11001263, the NKBRSFC under grants Nos. 2009CB930701,
2010CB922904, and 2011CB921502, and NSFC-RGC under grants Nos. 11061160490
and 1386-N-HKU748/10. D. S. Wang was supported by China Postdoctoral Science
Foundation.


\begin{thebibliography}{99}
\bibitem{Donnelly} R. J. Donnelly, Quantized Vortices in Helium II
(Cambridge University Press, Cambridge, 1991), Chaps. 4, 5.

\bibitem{Parkins} A. S. Parkins and D. F. Walls, Phys. Rep. 303, 1 (1998).

\bibitem{Pethick} C.J. Pethick and H. Smith, Bose-Einstein condensataion in
Dilute Gases, Cambridge University Press, Cambridge (2001).

\bibitem{Cooper} N. R. Cooper, Adv. in Phys., 57, 539 (2008).

\bibitem{Butts} D. A. Butts and D. S. Rokhsar, Nature (London) 397, 327
(1999).

\bibitem{Engels} P. Engels, I. Coddington, P. C. Haljan, V. Schweikhard, and
E. A. Cornell, Phys. Rev. Lett. 90, 170405 (2003).

\bibitem{Schweikhard} V. Schweikhard, I. Coddington, P. Engels, V. P.
Mogendorff, and E. A. Cornell, Phys. Rev. Lett. 92, 040404 (2004).

\bibitem{Kevrekidis2007} P. G. Kevrekidis, D. J. Frantzeskakis, Ricardo
Carretero-Gonzalez (Editors), Emergent Nonlinear Phenomena in Bose-Einstein
Condensates: Theory and Experiment, Springer, 2007.

\bibitem{Dalfovo} F. Dalfovo, S. Giorgini, L. Pitaevskii and S. Stringari,
Rev. Mod. Phys. 71, 463 (1999); D. L. Feder, A. A. Svidzinsky, A. L. Fetter
and C. W. Clark, Phys. Rev. Lett. 86, 564 (2001).

\bibitem{Aftalion_Ueda} J. J. Garc${\rm \acute{i}}$a-Ripoll and V.M. P${\rm \acute{e}}$ez-Garc${\rm \acute{i}}$a, Phys. Rev. A 64, 053611 (2001); A. Aftalion and R. L. Jerrard, Phys. Rev. A  66, 023611 (2002); H. Saito and M. Ueda, Phys. Rev. Lett. 93, 220402 (2004).

\bibitem{Rosenbusch} P. Rosenbusch, V. Bretin and J. Dalibard, Phys. Rev.
Lett. 89, 200403 (2002)

\bibitem{Penckwitt} A. A. Penckwitt, R. J. Ballagh and C. W. Gardiner, Phys.
Rev. Lett. 89, 260402 (2002).

\bibitem{Pu-PRA-1999} H. Pu, C. K. Law, J. H. Eberly, and N. P. Bigelow,
Phys. Rev. A 59, 1533 (1999); Y. Kawaguchi and T. Ohmi, Phys. Rev. A 70,
043610 (2004).

\bibitem{Jackson} A. D. Jackson, G. M. Kavoulakis, and E. Lundh, Phys. Rev. A
72, 053617 (2005); E. Lundh and H. M. Nilsen, Phys.
Rev. A 74, 063620 (2006).

\bibitem{JAMH} J. A. M. Huhtam\"{a}ki, M. M\"{o}tt\"{o}nen, and S. M. M.
Virtanen, Phys. Rev. A 74, 063619 (2006).


\bibitem{Kumakura-2007} M. Kumakura, and Y. Takahashi, Phys.
Rev. Lett. 99, 200403 (2007); H. M. Nilsen and E. Lundh, Phys. Rev. A 77,
013604 (2008); P. Kuopanportti and M. Mottonen, Phys. Rev. A 81, 033627
(2010).

\bibitem{Simula-PRA-2002} T. P. Simula, S. M. M. Virtanen, and M. M.
Salomaa, Phys. Rev. A 65, 033614 (2002).

\bibitem{Lundh-PRA-2002} E. Lundh, Phys. Rev. A 65, 043604 (2002); C.
Josserand, Chaos 14, 875 (2004); A. D. Jackson, G. M. Kavoulakis, and E.
Lundh, Phys. Rev. A 69, 053619 (2004); H. Fu and E. Zaremba, Phys. Rev. A
73, 013614 (2006).

\bibitem{Ketterle2001}J. R. Abo-Shaeer, C. Raman, J. M. Vogels, and W. Ketterle,
Science 292, 476 (2001).

\bibitem{Ruostekoski-PRA-2004} J. Ruostekoski, Phys. Rev. A 70, 041601(R)
(2004).

\bibitem{Sulem} C. Sulem and P. Sulem, The Nonlinear Schrodinger Equation
Springer-Verlag, Berlin, 1999.

\bibitem{Belmonte-Beitia1} J. Belmonte-Beitia, V. M. Pereez-Garcia and V.
Vekslerchik, Phys. Rev. Lett. 98, 064102 (2007);
M. Salerno, V. V. Konotop, and Yu. V. Bludov, Phys. Rev. Lett. 101, 030405 (2008).

\bibitem{Wang} D. S. Wang, X. H. Hu, J. Hu and W. M. Liu, Phys. Rev. A 81,
025604 (2010).

\bibitem{Belmonte-Beitia2} J. Belmonte-Beitia, V. M. Pereez-Garcia,
V.Vekslerchik and V. V. Konotop, Phys. Rev. Lett. 100, 164102 (2008); D. S.
Wang, X. H Hu and W. M. Liu, Phys. Rev. A 82, 023612 (2010).

\bibitem{Wu} L. Wu, L. Li, J. F. Zhang, D. Mihalache, B. A. Malomed and W.
M. Liu, Phys. Rev. A 81, 061805(R) (2010).

\bibitem{Yan_Konotop} Z. Yan and V. V. Konotop, Phys. Rev. E 80, 036607
(2009); Z. Yan, V. V. Konotop and N. Akhmediev, Phys. Rev. E 82, 036610 (2010).

\bibitem{Kasamatsu} K. Kasamatsu, M. Tsubota and M. Ueda, Phys. Rev. A 66,
053606 (2002).

\bibitem{Timmermans} E. Timmermans, P. Tommasini, M. Hussein and A. Kerman,
Phys. Rep. 315, 199 (1999).

\bibitem{Roberts} J. L. Roberts, N. R. Claussen, J. P. Burke, C. H. Greene,
E. A. Cornell and C. E. Wieman, Phys. Rev. Lett. 81, 5109 (1998); E. A.
Donley, N. R. Claussen, S. L. Cornish, J. L. Roberts, E. A. Cornell and C.
E. Wieman, Nature 412, 295 (2001).

\bibitem{Mihalache} D. Mihalache, D. Mazilu, B. A.Malomed and F. Lederer,
Phys. Rev. A 73, 043615 (2006); L. D. Carr and C. W. Clark, Phys. Rev. Lett.
97, 010403 (2006).

\bibitem{Saito_Ueda} H. Saito and M. Ueda, Phys. Rev. Lett. 90, 040403
(2003).

\bibitem{Abramowitz} M. Abramowitz and I. A. Stegun, (Eds.). Confluent
Hypergeometric Functions, New York: Dover, 503-515 (1972).

\bibitem{Cornish} S. L. Cornish, N. R. Claussen, J. L. Roberts, E. A.
Cornell and C. E. Wieman, Phys. Rev. Lett. 85, 1795 (2000).

\bibitem{Bradley} C. C. Bradley, C. A. Sackett, J. J. Tollett and R. G.
Hulet, Phys. Rev. Lett. 75, 1687 (1995).

\bibitem{Gerton} J. M. Gerton, D. Strekalov, I. Prodan and R. G. Hulet,
Nature 408, 692 (2000); S. E. Pollack, D. Dries, M. Junker, Y. P.
Chen, T. A. Corcovilos and R. G. Hulet, Phys. Rev. Lett. 102, 090402 (2009).

\bibitem{Anderson} M. H. Anderson, J. R. Ensher, M. R. Matthews, C. E.
Wieman and E. A. Cornell, Science 269, 198 (1995); S. Burger, K. Bongs, S.
Dettmer, W. Ertmer and K. Sengstock, Phys. Rev. Lett. 83, 5198 (1999).

\bibitem{Moerdijk} A. J. Moerdijk, B. J. Verhaar and A. Axelsson, Phys. Rev.
A 51, 4852 (1995).

\bibitem{Theocharis} G. Theocharis, D. J. Frantzeskakis, P. G. Kevrekidis,
B. A. Malomed and Y. S. Kivshar, Phys. Rev. Lett. 90, 120403 (2003).

\bibitem{Ueda_Liu}H. Saito and M. Ueda, Phys. Rev. A 70, 053610 (2004);
C. N. Liu, T. Morishita and S. Watanabe, Phys. Rev. A 75, 023604 (2007).

\bibitem{Madison2000} K. W. Madison, F. Chevy, W. Wohlleben, and J. Dalibard,
 Phys. Rev. Lett. 84, 806 (2000).

\bibitem{Rychtarik} D. Rychtarik et al., Phys. Rev. Lett. 92, 173003
(2004).

\end{thebibliography}
\end{document}